\documentclass[11pt]{amsart}
\usepackage{epsfig}

\usepackage{listings} 

\usepackage{amscd}
\usepackage{amsmath}
\usepackage{latexsym}
\usepackage{amsfonts}
\usepackage{amssymb}

\theoremstyle{plain}

\numberwithin{equation}{section}

\newcommand\bes{\begin{eqnarray}}
\newcommand\ees{\end{eqnarray}}
\newcommand\bess{\begin{eqnarray*}}
\newcommand\eess{\end{eqnarray*}}

\begin{document}
\setlength\baselineskip{16pt}
\title[ ]
{A Steady State Solution to a Mortgage Pricing Problem}

\author{Dejun Xie}

\address{Department of Mathematics,
University of Delaware, Newark, DE 19711, $dxie@math.udel.edu$}

\begin{abstract}
This paper considers a mortgage contract where the borrower pays a
fixed mortgage rate and has the choice of making prepayment.
 Assume the market interest follows the CIR model, the problem is formulated as
 a free boundary problem where the free boundary denotes the level of market interest rate
 at which it is optimal for the borrower to make prepayment.  Here
 we focus on the infinite horizon problem.  Using variational method,
 we obtain an analytical solution to the problem, where the free boundary is implicitly given by a transcendental algebraic equation.
\end{abstract}

\maketitle

\section{Formulation of the problem}

We consider a mortgage contract where the borrower pays a fixed
rate of $c$ $(year^{-1})$ to the lender. In reality this mortgage
rate is implicitly represented by a continuous payment of $m$ $
\$/year$.  At each time $t$ when the contract is effect, the
borrower has two choices: to continue the mortgage by paying $mdt$
for the next $dt$ period or to close the mortgage by paying off
all the loan balance $M(t)$, where the loan balance $M(t)$ is
determined by \bes \label{ODEforM}\frac{dM(t)}{dt}=-m+cM(t).\ees
When the contract duration $T$ is given and $M(T)=0$ specified,
the above ODE has a unique soultion \bes
M(t)=\frac{m}{c}(1-e^{(-c(T-t))}).\ees Here we assume the borrower
always has sufficient amount of capital. The borrower chooses not
to pay $M(t)$ even though he is financially capable to do so if
the expected future market return from an equal amount of
investment is higher enough. On the other hand, if the expected
future market return from an equal amount of investment is lower
enough, he should choose to settle $M(t)$.  From the lender's
point of view, the value of the contract $V$, as a function of
time $t$ and market interest rate $x$, is determined by the market
interest rate $x$. The higher the market interest rate $x$ is, the
lower the contract value $V$, but $V$ shall never be lower than 0.
The lower the market interest rate $x$ is, the higher the contract
value $V$, but $V$ shall never be exceed $M(t)$ since the borrower
has the choice to settle the loan once $V$ reaches $M(t)$. From
standard mathematical finance theory, one can find the value of
the contract $V(x,t)$ and the optimal level of market interest
$x=h(t)$ at which the borrower should make prepayment of $M(t)$ by
solving the following free boundary problem: \bes
\label{orignalSystem} \left\{\begin{array}{ll}\displaystyle
\textbf{L}(V)=m ,\quad & $for$\quad x> h(t) , t>0
\medskip\\ \displaystyle V=\frac{m}{c}[1-e_{{}}^{-ct}] ,\quad &
$for$\quad x\leq h(t) ,t>0 \medskip\\ \displaystyle \frac{\partial
V}{\partial x}(h(t),t) \equiv 0  \medskip \\  V(x,0)=0, \quad &
$for all$\quad x\geq 0 \medskip  \\
h(0)=c
\end{array}\right.
 \ees
 where $t$, for mathematical convenience, is defined to be the
 time to expiry of the contract, $h(t)$ is the unknown free boudary to be determined together with
$V$, and the differential operator  $\textbf{L}$ is defined as
 \bes \label{Operator}\textbf{L}(V)=\frac{\partial V}{\partial t} - \frac{\sigma^2}2 x \frac{\partial^2
V}{\partial x^2} - k(\theta- x) \frac{\partial V}{\partial x} + x
V \ees

%
%
%
The differential operator in the the system, referred as Kolmogorov
equation, which can also be derived from Feymann-Kac Theorem
\cite{Wilmont, Hull, Rogers}. Because of the important role played
by the mortgage securities in real economy, there exists a
considerable literature (see \cite{Baxter, BuserMortgage, EKKM,
EppersonMortgage, KauMortgage}, for instance) dedicated to the
topic, mostly of which have studied the problem from
option-theoretical point with relatively less rigorous mathematical
proofs.  In recent development, a free boundary approach was
introduced in \cite{Xie, JiangMortgage} to a similar problem. In
particular D. Xie et al. (2007) have formulated the integral
representations of the problems and proposed a Newton's iteration
scheme under the assumption that the underlying interest rate
follows the Vasicek model \cite{Xie}. In this paper, we shall study
the same type of mortgage contract with market interest rate
following CIR model instead of Vasicek model.  We use CIR model
because it is observed that Vasicek model allows negative interest
rate, which contradicts the empirical statistics from market
\cite{CIRmodel}.  In this paper, we focus on the infinite horizon
(steady state) of the problem.

\section{The Infinite Horizon Problem}
Without loss of generality, we assume $m=c$, thus we derive the
following infinite horizon problem of the original system. \bes
\left \{ \begin{array} {ll} \displaystyle
xV^{\prime\prime}+(\frac{2k\theta}{\sigma^{2}}-\frac{2k}{\sigma^{2}%
}x)V^{\prime}-\frac{2}{\sigma^{2}}xV=-\frac{2c}{\sigma^{2}}, \qquad x \in (R^{\ast},~\ \infty) \\
V(R^{\ast})=1 \\
V_{x}(R^{\ast})=0\\
V(x=\infty)=0
\end{array} \right. \label{InfiniteHorizonSystem} \ees
where $R^{\ast\text{ }}$ is to be determined together with $V.$ To
solve this free boundary problem, we let \bess V(x)&=&e^{\lambda
x}u(z)  \\z&=&px=\frac{2\sqrt{k^{2}+2\sigma^{2}}}{\sigma^{2}}x
\\\lambda&=&\frac{k-\sqrt{k^{2}+2\sigma^{2}}}{\sigma^{2}} \eess
The above infinite horizon problem is transformed into the
following one. \bess \left \{ \begin{array} {ll} \displaystyle
zu^{\prime\prime}+(\frac{2k\theta}{\sigma^{2}}-z)u^{\prime}-\frac{k\theta
}{\sigma^{2}}(1-\frac{k}{\sqrt{k^{2}+2\sigma^{2}}})u =
-\frac{c}{\sqrt
{k^{2}+2\sigma^{2}}}e^{(\frac{1}{2}-\frac{k}{2\sqrt{k^{2}+2\sigma^{2}}})z},
  z \in (z^{\ast},\infty)  \\
u(z^{\ast})=e^{(\frac{1}{2}-\frac{k}{2\sqrt{k^{2}+2\sigma^{2}}})z^{\ast}}
\\
u^{\prime}(z^{\ast}) =
(\frac{1}{2}-\frac{k}{2\sqrt{k^{2}+2\sigma^{2}}
})e^{(\frac{1}{2}-\frac{k}{2\sqrt{k^{2}+2\sigma^{2}}})z^{\ast}}
\\
e^{\frac{\lambda}{p}z}u(z)=0, \qquad z \rightarrow \infty
\end{array} \right. \eess We first solve the ODE in $z$ , disregarding any
boundary conditions. \ The homogeneous equation, as a standard
confluent hypergeometric equation, has two linearly independent
solutions \bess
u_1=M(\frac{k\theta}{\sigma^{2}}(1-\frac{k}{\sqrt{k^{2}+2\sigma^{2}}}%
),\frac{2k\theta}{\sigma^{2}},z) \\
u_2=U(\frac{k\theta}{\sigma^{2}}(1-\frac
{k}{\sqrt{k^{2}+2\sigma^{2}}}),\frac{2k\theta}{\sigma^{2}},z)\eess
one condition that both $\frac{k\theta}{\sigma^{2}}(1-\frac{k}{\sqrt
{k^{2}+2\sigma^{2}}})$ and $\frac{2k\theta}{\sigma^{2}}$ are
strictly positive, which is certainly satisfied for this problem.
And the Wronskian of these two linearly independent solutions is
calculated to be \bess
W(M,U)(z)=-\frac{\Gamma(\frac{2k\theta}{\sigma^{2}})}{\Gamma(\frac{k\theta
}{\sigma^{2}}(1-\frac{k}{\sqrt{k^{2}+2\sigma^{2}}}))}z^{-\frac{2k\theta
}{\sigma^{2}}}e^{z}. \eess By the standard variational method, we
can find one particular solution to the inhomogeneous equation, and
together with above two linearly independent solutions, we obtain
the general solution of the ODE for $u$ \bess \begin{split} & u(z)=c_{1}\ M(\alpha,\gamma,z)+c_{2}U(\alpha,\gamma,z) \\
&  \quad +\int_{z}^{\infty}%
\frac{M(\alpha,\gamma,\xi)U(\alpha,\gamma,z)-U(\alpha,\gamma,\xi
)M(\alpha,\gamma,z)}{-\frac{\Gamma(\gamma)}{\Gamma(\alpha)}\xi^{-\gamma}%
}(-\frac{c}{\sqrt{k^{2}+2\sigma^{2}}})e^{-(\frac{1}{2}+\frac{k}{2\sqrt
{k^{2}+2\sigma^{2}}})\xi}d\xi  \end{split},\eess and
correspondingly, the general solution to the ODE in
(\ref{InfiniteHorizonSystem}) is given by \bes
\begin{split}  & V(x)=e^{\lambda
x}\{c_{1}\ M(\alpha,\gamma,px)+c_{2}U(\alpha,\gamma,px) \\
& \quad
-\int_{z}^{\infty}\frac{M(\alpha,\gamma,\xi)U(\alpha,\gamma,px)-U(\alpha
,\gamma,\xi)M(\alpha,\gamma,px)}{-\frac{\Gamma(\gamma)}{\Gamma(\alpha)}%
\xi^{-\gamma}}[\frac{ce^{-(\frac{1}{2}+\frac
{k}{2\sqrt{k^{2}+2\sigma^{2}}})\xi}}{\sqrt{k^{2}+2\sigma^{2}}}]d\xi\}
\end{split} \ees
where \bess \alpha &=&
\frac{k\theta}{\sigma^{2}}(1-\frac{k}{\sqrt{k^{2}+2\sigma^{2}}})\\
\gamma &=& \frac{2k\theta}{\sigma^{2}}\\
p&=&\frac{2\sqrt{k^{2}+2\sigma^{2}}}{\sigma^{2}}\eess

Next we need to decide, by the conditions at free boundary and
infinity, the values of two constants in the general solution as
well as the unknown infinite horizon z$^{\ast}.$ We firstly convince
ourselves that $c_{1}=0$ by investigation of the asymptotic behavior
of the $\ M(\alpha,\gamma,z)$ and $U(\alpha,\gamma,z)$ for large z.
Recall that \bess M(\alpha,\gamma,z) &\sim&
\frac{\Gamma(\gamma)}{\Gamma(\alpha )}z^{\alpha-\gamma}e^{z}, \quad
z \rightarrow +\infty \\
U(\alpha,\gamma,z) &\sim& x^{-\alpha}, \quad z \rightarrow +\infty.
\eess If we let $z \rightarrow \infty$ in the expression of $u(z)$,
the definite integral term vanishes, i.e., the $U$ part contribution
goes zero, but the $M$ part go to $e^{z}.$ If we multiply them with
$e^{\frac{\lambda}{p}z},$ the magnitude contributed by the definite
integral or the $U$ part still go zero, which is the desired,
because $\lambda<0.$ \ But now the contribution of $M$ will be $
e^{\frac{\lambda}{p}z}e^{z}$ multiplying a nontrivial polynomial in
$z$, which clearly goes to infinity. This implies that
$M(\alpha,\gamma,z)$ should not be included in the general solution.
Now we only have two unknowns $c_{1\text{ }}$and $z^{\ast}.$ It is
easy to use the fact that \bess
u(z^{\ast})&=&e^{(\frac{1}{2}-\frac{k}{2\sqrt
{k^{2}+2\sigma^{2}}})z^{\ast}}\\ u^{\prime}(z^{\ast})&=&(\frac{1}{2}%
-\frac{k}{2\sqrt{k^{2}+2\sigma^{2}}})e^{(\frac{1}{2}-\frac{k}{2\sqrt
{k^{2}+2\sigma^{2}}})z^{\ast}}\eess to get the implicit solution of
$z^{\ast}\,:$
\bess  \displaystyle \begin{split} & \frac{U(\alpha,\gamma,z^{\ast})}{U(\alpha+,\gamma+1,z^{\ast})}%
=
\\ & \quad \frac{
e^{(\frac{1}{2}-\frac{k}{2\sqrt{k^{2}+2\sigma^{2}}})z^{\ast}}+ 
\frac
{c}{\sqrt{k^{2}+2\sigma^{2}}}\frac{\Gamma^2(\alpha)}{\Gamma^2(\gamma)}%
\int_{z^{\ast}}^{\infty}\frac{M(\alpha,\gamma,\xi)U(\alpha,\gamma,z^{\ast
})-U(\alpha,\gamma,\xi)M(\alpha,\gamma,z^{\ast})}{\xi^{-\gamma} e^{(\frac{1}{2}+\frac{k}{2\sqrt{k^{2}+2\sigma^{2}}})\xi}}d\xi}{e^{(\frac{1}{2}-\frac{k}{2\sqrt{k^{2}+2\sigma^{2}}%
})z^{\ast}}+
\frac{c}{\sqrt{k^{2}+2\sigma^{2}}}\frac{\Gamma^2(\alpha)}%
{\Gamma^2(\gamma)} 
\int_{z^{\ast}}^{\infty}\frac{M(\alpha,\gamma,\xi
)U(\alpha+1,\gamma+1,z^{\ast})-U(\alpha,\gamma,\xi)M(\alpha+1,\gamma
+1,z^{\ast})}{\xi^{-\gamma}e^{(\frac{1}{2}+\frac{k}{2\sqrt{k^{2}+2\sigma^{2}}})\xi}}%
d\xi}
\end{split} \eess
And the $x^{\ast}$ in original system (\ref{orignalSystem}) can be
simply recovered by $x^{\ast}=pz^{\ast }.$


\begin{thebibliography}{999}



%


\bibitem{CIRmodel} Cox, John, Ingersoll, Jonathan, \& Ross, Stephen, {\em A theory of
the term structure of interest rates}, Econometrica, {\bf
53}(1985), 385-407.

%
%


\bibitem{XinfuConvexity} X. Chen, J. Chadam , L. Jiang \& W. Zhang, { Convexity of the exercise boundary
of the American put option on a zero dividend asset}, Mathematical
Finance {\bf 18} (2008), 185-197.


\bibitem{Hull} Hull, J. C. \& A. White  , {\em Pricing interest rate derivative
securities}, Review of Financial Studies, {\bf 3} (1990) 573-592 .

\bibitem{XinfuAmericanPut} Xinfu Chen \& J. Chadam, {Mathematical analysis of an American put option},
 SIAM J. Math. Anal. {\bf 38} (2007) 1613--1641.


\bibitem{Xie} Xie, Dejun, Chen, Xinfu \& Chadam, John, {\em Optimal Termination of Mortgages}, European Journal of
Applied Mathematics, {\bf 3}(2007), 363-388.

%

\bibitem{Baxter} M. Baxter \& A. Rennie, {\em Financial Calculus, An
introductin to derivative pricing}, Cambridge University Press,
2005.


\bibitem{BuserMortgage} S.A.  Buser, \& P. H. Hendershott, {\em Pricing
default-free fixed rate mortgages}, Housing Finance Rev. {\bf 3}
(1984), 405--429.




\bibitem{EppersonMortgage} J. Epperson, J.B. Kau, , D.C. Keenan,  \& W. J. Muller,
{\em Pricing default risk in mortgages}, AREUEA J. {\bf 13}
(1985), 152--167.

\bibitem{Friedman} A. Friedman, {\em Variational Principles and Free
Boundary Problems,} John Wiley \& Sons, Inc., New York, 1982.


\bibitem{JiangMortgage} L. Jiang, B. Bian \& F. Yi.
{\em A parabolic variational inequality arising from the valuation
              of fixed rate mortgages},
European J. Appl. Math. {\bf 16} (2005), 361--338.


\bibitem{KauMortgage} J.B.  Kau \& D.C.  Keenan,  {\em An Overview of the
option-theoretic pricing of mortgages},  J. Housing Res. {\bf 6}
(1995), 217--244.


\bibitem{StantonMorgageChoice} Stanton, R., \& N. Wallace, {\em Mortgage choice: what's the
point? }, Real Estate Economics, {\bf 26} (1984), 173-205.



\bibitem{Wilmont} P. Willmott, {\em Derivatives, the theory and
practice of financial engineering}, John Wiley \& Sons, New York,
1999.


\bibitem{Merton} R. Merton, {\em On the pricing of corporate debt: the risk of structure of interest rates},
Journal of Finance, {\bf 29} (1974), 449--469.




\bibitem{Jeffrey} A. Jeffrey \& D. Zwillinger, {\em Table of Integrrals, Series, and Products
}, 6th Edition, 890-892.

\bibitem{AtlasOfFunctions} J. Spanier \& K. B. Oldham, {\em An Atlas of Functions}, 436-443 .

\bibitem{Bossavit}  A. Bossavit, A. Damlamian , \& M. Fremond,
{\em Free Boundary Problems: Applications and Theory  }, Pitman
Pub Ltd, 1986.



\bibitem{BR} M. Baxter \& A. Rennie, {\sc Financial Calculus, An
introductin to derivative pricing}, Cambridge University Press,
2005.

\bibitem{EKKM} J. Epperson, J.B. Kau, , D.C. Keenan,  \& W. J. Muller,
{\em Pricing default risk in mortgages}, AREUEA J. {\bf 13}
(1985), 152--167.

%


\bibitem{HullBook} J. Hull, {\em Options, Futures and Other Derivatives
},
 Prentice Hall,
2005.


\bibitem{Neft} S. N. Neftci , {\em An Introduction to the Mathematics of Financial Derivatives
},
 Academic Press,
2000.

\bibitem{Allison} A. Etheridge, {\em A course in Financial Calculus},
 Cambridge University Press,
2004.




\bibitem{Rogers} L. C. Rogers \& D. Talay {\em Numerical Methods in Finance}, Cambridge University Press,
2007.

\bibitem{NumericalFinance2} J. Miller {\sc Numerical Methods in Finance}, CRC
Pr I Llc, 2007.


\end{thebibliography}
\end{document}